\documentclass[11pt]{article}
\usepackage{array}
\usepackage{graphicx}
\usepackage{cite}
\usepackage{textpos}

\title{Unification of Force and Substance\footnote{Talk at Royal Society Symposium, ``Unifying Physics and Technology in the Light of Maxwell's Equations'', November 2015.}}
\author{Frank Wilczek \\
\small\it Center for Theoretical Physics, MIT, Cambridge MA 02139 USA}

\begin{document}

\maketitle

\begin{textblock*}{5cm}(11cm,-8.2cm)
\fbox{\footnotesize MIT-CTP-4744}
\end{textblock*}

\begin{abstract}
Maxwell's mature presentation of his equations emphasized the unity of electromagnetism and mechanics, subsuming both as ``dynamical systems''.   That intuition of unity has proved both fruitful, as a source of pregnant concepts, and broadly inspiring.  A deep aspect of Maxwell's work is its use of redundant potentials, and the associated requirement of gauge symmetry.  Those concepts have become central to our present understanding of fundamental physics, but they can appear to be rather formal and esoteric.  Here I discuss two things: The physical significance of gauge invariance, in broad terms; and some tantalizing prospects for further unification, building on that concept, that are visible on the horizon today.   If those prospects are realized, Maxwell's vision of the unity of field and substance will be brought to a new level.   
\end{abstract}

\medskip

As we celebrate the sesquicentennial of Maxwell's equations, there are many facets to celebrate: the triumph of the field concept, the roots of relativity, the cornerstone of a host of technologies.  In many ways, the Maxwell equations mark the beginning of modern physics, and indeed of the modern world.   

Here I will discuss a special, hidden theme of Maxwell's work, primarily important for fundamental physics, whose profundity only emerged gradually, and may still not stand fully revealed even today.   This is the concept of gauge symmetry.   Gauge symmetry is perhaps the supreme realization of Heinrich Hertz's famous tribute to the Maxwell equations:
\begin{quote}
One cannot escape the feeling that these mathematical formulae have an independent existence and an intelligence of their own, that they are wiser than we are, wiser even than their discoverers, that we get more out of them than was originally put into them.
\end{quote}

\section{Gauge Symmetry}

{\it Local gauge symmetry\/} arises when one formulates the Maxwell equations using potentials.  It states that one can subject the potentials to an enormous group of transformations, parameterized by a function of space and time, while leaving their physical consequences unchanged.   On the surface this seems like a very odd principle.  It begs the question, why one uses the (largely arbitrary) potentials at all, rather than working in terms of invariants. Yet this principle, suitably generalized, has proved to be a reliable guide to formulating successful laws of fundamental physics.  In different forms, it is central to our theories of the strong, weak, and gravitational interactions as well as electromagnetism.  It may point the way to more profound, unified understanding of all those forces, as I shall discuss.   

How did Maxwell get to this concept?  Why does Nature use it?  Those are the questions I'd like to address now.

\subsection{Maxwell, Faraday, and Hamilton}

Maxwell's mature presentation of his equations, in the {\it Treatise} \cite{treatise}, emphasized the unity of electromagnetism and mechanics, subsuming both in the concept of (Hamiltonian) ``dynamical system''.   In Article \S 567 , he wrote
\begin{quote}
In formulating the ideas and words relating to any science, which, like electricity, deals with forces and their effects, we must keep constantly in mind the ideas appropriate to the fundamental science of dynamics, so that we may, during the first development of the science, avoid inconsistency with what is already established ...
\end{quote}

At the same time, Maxwell was determined to embody the physical intuitions of Faraday.  In the preface to the first edition of the {\it Treatise\/} he wrote, contrasting Faraday's conceptions to the action-at-a-distance theories then more popular among mathematical physicists
\begin{quote}
... Faraday, in his mind's eye, saw lines of force traversing all space where the mathematicians saw centres of force attracting at a distance: Faraday saw a medium where they saw nothing but distance: Faraday sought the seat of the phenomena in real actions going on in the medium, they were satisfied that they had found it in a power of action at a distance impressed on the electric fluids.
\end{quote}
and a paragraph later
\begin{quote}
The whole theory, for instance, of the potential, considered as a quantity which satisfies a certain partial differential equation, belongs essentially to the method which I have called that of Faraday.
\end{quote}

Faraday's concept of ``electrotonic state'' was a non-mathematical anticipation of the vector potential, which Maxwell took over in formulating his equations.   For Faraday, it was a way of expressing how his space-filling fields influence electric currents.  It is perfectly adapted to Maxwell's program of founding the description of electromagnetic phenomena on the mechanical notion of local energy density, since it allows one to express the energy density $\vec A \cdot \vec j$  associated with a current $\vec j$.

The arbitrariness of the potentials is reflected in the possibility that one can make a wide class of transformations upon them, without changing their physical content.   The fields
\begin{eqnarray}
\vec E ~&=&~ - \vec \nabla \phi \, - \, \frac{\partial {\vec A}}{\partial t} \nonumber \\
\vec B ~&=&~ \vec \nabla \times \vec A
\end{eqnarray}
are manifestly invariant under the transformations
\begin{eqnarray}\label{gaugeTransformation}
\phi^\prime ~&=&~ \phi \, + \, \frac{\partial\Lambda}{\partial t} \nonumber \\
{\vec A}^\prime ~&=&~ \vec A - \vec \nabla \Lambda
\end{eqnarray}
The invariance of the work expressions is less obvious.  We have
\begin{eqnarray}
&{}& \ \rho \phi^\prime \, - \, \vec j \cdot {\vec A}^\prime \nonumber \\
~&=& \rho \phi \, - \, \vec j \cdot {\vec A} \, + \, \rho \frac{\partial \Lambda}{\partial t}  \, + \, \vec j \cdot \vec \nabla \Lambda \nonumber \\ 
~&=&~  \rho \phi \, - \, \vec j \cdot {\vec A} \,  + \, \{ \partial_t (\rho \Lambda) + \nabla (\vec A \cdot \vec j) \} \, - \, \{ \{ \Lambda (\partial_t \rho + \vec \nabla \cdot \vec j ) \} \} \nonumber  \\
~&\approx&~ \rho \phi \, - \, \vec j \cdot {\vec A} 
\end{eqnarray}
where the term in single braces vanishes upon integration, as long as $\Lambda$ vanishes at infinity.  The term in double braces vanishes if, and in general only if, charge is locally conserved.  

The relationship between gauge invariance and charge conservation becomes especially transparent in relativistic notation.  The invariance of the coupling (energy) expression $\int A_\mu j^\mu$ under $A_\mu \rightarrow A_\mu + \partial_\mu \Lambda$ is equivalent, upon integration by parts, to the conservation equation $\partial_\mu j^\mu =0$.

In keeping with my theme of unification, it is appropriate to add that Hamilton was led to his formulation of mechanics in large part by a desire to unite the mathematical treatment of mechanics and optics, including wave optics.  Following the experimental discoveries of quantum behavior, those alternative descriptions enabled a smooth implementation of wave-particle duality.  Indeed, the mathematical blueprint for passing from Hamiltonian ``wave'' mechanics, epitomized in the Hamilton-Jacobi equation, to the Schr\"odinger equation is astoundingly simple.

\subsection{Quantum Theory}

\subsubsection{Dirac}

Dirac spelled out the abstract ``particle'' formulation of Hamiltonian dynamics for electromagnetism, which Maxwell had gestured toward (but did not work out in detail).   In that formulation, the modes of the electromagnetic field appear as independent harmonic oscillators.  Dirac could then apply his general quantization procedure, replacing Poisson brackets by commutators, to get a quantum theory of the electromagnetic field \cite{diracEM}.   This theory has proved to be tremendously accurate.  It has been supplemented, but not replaced, within our current best understanding of Nature.

The quantum theory, through its {\it reliance\/} on potentials, sheds new light on a basic question which the Maxwell equations suggest.
The is no difficulty in introducing magnetic charges and currents into the classical Maxwell equations, at the level of field strengths.  The system of equations including magnetic charges and currents
\begin{eqnarray}
\vec \nabla \cdot \vec B ~&=&~ \rho_m \nonumber \\
\vec \nabla \cdot \vec E ~&=&~ \rho \nonumber \\
\vec \nabla \times \vec B ~&=&~ \frac{\partial \vec E}{\partial t} + \vec J \nonumber \\
\vec \nabla \times \vec E ~&=&~ -\frac{\partial \vec B}{\partial t} - \vec J_m
\end{eqnarray}
supports a conservation law for magnetic charge
\begin{equation}
\frac{\partial \rho_m}{\partial t} + \vec \nabla \cdot \vec J_m ~=~ 0
\end{equation}
and is no less consistent than the original Maxwell equations.  It even appears more symmetric.   
From this perspective, the observed absence of free magnetic charge seems puzzling. Why doesn't Nature make use of that possibility?   

In quantum theory, however, the picture is different.   Quantum theory, in all known formulations, requires not only field strengths, but also vector potentials.    Once we introduce vector potentials, to write field strengths in the form (now using relativistic notation) 
\begin{equation}
F_{\alpha \beta} ~=~ \partial_\alpha A_\beta - \partial_\beta A_\alpha 
\end{equation}
with
\begin{eqnarray}
\vec E_j ~&=&~ - F_{0j} \\
\vec B_j ~&=&~ \frac{1}{2} \epsilon_{jkl} F_{kl}
\end{eqnarray}
we find that the first Maxwell equation in its usual form, stating the absence of magnetic charge
\begin{equation}
\vec \nabla \cdot \vec B ~=~ 0
\end{equation}
holds true as an identity, as does Faraday's law of induction.   

It is remarkable that Faraday's great empirical discovery, which came as a great surprise in its time, now appears as a mathematical identity, required by the logic of profound theoretical principles.   

Dirac later demonstrated \cite{diracMonopole} that if we loosen the framework, by allowing certain kinds of singularities in the vector potentials, then we can introduce magnetic charge while maintaining many of the basic principles of relativistic quantum theory.    There are significant -- possibly insuperable -- difficulties in accommodating such singularities within a fully dynamical theory of electrodynamics.   But in a final twist to this story, 'tHooft \cite{'tHooft} and Polyakov \cite{polyakov} showed that if electrodynamics is embedded within larger gauge theories, including the possibly realistic ones described below, then magnetic monopoles can appear without requiring singular potentials.  The extra degrees of freedom that resolve the singularity of the purely electromagnetic potentials are predicted to be very massive, so the magnetic monopoles are predicted to be very heavy -- out of the reach of accelerators\footnote{The absence of residual magnetic monopoles emerging from the big bang is a separate issue.  Their observed absence was one of the original motivations for introducing the idea of cosmic inflation.}.  From a larger perspective, it is a major achievement of Hamiltonian mechanics, and of the quantum theory that builds upon it, to have explained the pronounced {\it asymmetry\/} between electric and magnetic charge, by emphasizing the need to introduce potentials.

\subsubsection{Longitudinal Modes}

Given that potentials are desirable, in order to formulate electromagnetic field theory as a dynamical system, we may ask why Nature chooses to make them largely redundant, by enforcing gauge symmetry.  Relativistic quantum theory supplies a profound answer to that question.  

Gauge transformations, as shown in Eqn.\,(\ref{gaugeTransformation}), allow us to change the longitudinal (in the 3+1 dimensional sense) part of the potential field at will.   Gauge symmetry, therefore, basically states that the longitudinal modes have no direct physical significance.  They appear in the equations, but not in the consequences of those equations.   In common jargon, we say that longitudinal photons decouple.   

That decoupling of longitudinal photons is an important consistency requirement, because special relativity forces a relativistically invariant norm to have opposite sign for photons with timelike versus spacelike polarization.   If we are going to identify the norm with a probability, we must make sure that the timelike photons do not leak into our world.  Gauge invariance insures this, as I'll now indicate more precisely and mathematically. 

Using local charge conservation, we can decompose the electric current at spatial momentum $\vec k$ into longitudinal and transverse parts, according to
\begin{equation}
\vec j ~=~ - \omega \frac{\vec k}{{\vec k}^2} \rho \, + \, {\vec j}^\perp
\end{equation}
Then we have 
\begin{equation}
\rho_1 \rho_2 \, - \, {\vec j}_1 \cdot {\vec j}_2 ~=~ \rho_1 \rho_2 (1 - \frac{\omega^2}{{\vec k}^2} )\, -  \, {\vec j}_1^\perp \cdot {\vec j}_2^\perp
\end{equation}
and so for the photon-mediated coupling, allowing all four polarizations,
\begin{equation}
\frac{\rho_1 \rho_2 \, - \, {\vec j}^\perp_1 \cdot {\vec j}^\perp_2}{\omega^2 - {\vec k}^2} ~=~ - \frac{\rho_1 \rho_2}{{\vec k}^2} \, - \, \frac{ {\vec j}^\perp_1 \cdot {\vec j}^\perp_2}{\omega^2 - {\vec k}^2}
\end{equation}
The two surviving terms are an instantaneous Coulomb force and a coupling to {\it transverse\/} dynamical radiation -- or, in the quantum theory, to transverse photons -- only.    Those effects form the core of physical electrodynamics.

\subsubsection{Working Backwards}

As the preceding manipulations indicate, one can work backwards, ``deriving'' the Maxwell equations as a consequence of gauge symmetry and special relativity.  (And as we've seen, gauge invariance itself is closely tied to the possibility of formulating a consistent quantum theory.)  That cluster of ideas remains valid, and becomes more precise and rigorous, in the formal treatment of relativistic quantum field theory.   One finds that gauge invariance and special relativity, together with general principles of locality and renormalizability, lead one directly and uniquely to modern quantum electrodynamics, with Maxwell's equations as its heart.

\subsection{Local Symmetry}

Gauge invariance is a vast symmetry.  It allows transformations that depend freely on time and space.  We say that such transformations define a local symmetry group.    

Similarly to how Maxwell's electrodynamics can be derived from local gauge invariance, Einstein's general relativity can be derived from symmetry under local -- that is, space-time dependent -- Lorentz transformations.    However, that understanding of general relativity -- and of electrodynamics -- reverses the historical order of discovery.    Einstein, guided by experimental facts  (equivalence principle, Newtonian limit)  arrived at general relativity through a mixture of geometrical ideas and inspired intuition \cite{einstein}.  Only later did Weyl \cite{weyl} clarify and emphasize the local symmetry aspect, in an attempt to understand both gravity and electrodynamics on a similar footing.   (See below.) 

On the other hand Yang and Mills \cite{yangMills}, in their pioneering attempt to generalize Maxwell electrodynamics, consciously started from local symmetry.  They showed how one could construct interesting relativistic field theories accommodating more complicated gauge transformations, where the parameter $\Lambda(x)$ is not simply a numerical function, but a function taking values in a compact group.  Amazingly, their mathematical construction forms the core of our present theories of the weak and strong interaction.

Though in fundamental physics the central significance of local symmetry is a relatively recent discovery, artists in many cultures have employed it for millennia, especially in decorative art.   In such art \cite{decorativeArt}, one often finds  large-scale patterns built up from symmetric shapes, such as equilateral triangles, squares, and circles.   The symmetry of such objects arises from the fact that we can rotate them through appropriate angles around their centers, without changing their overall form.  But of course each object in the pattern has its own distinct center, and so to realize the full symmetry we must allow transformations of space that vary from place to place.   Many design patterns also employ transformations of color, as well as shape, whose symmetry is local.

\subsection{Weyl's Gauge Symmetry}

\subsubsection{Gauge Symmetry as Geometry}

The historical origin of the term ``gauge'' in gauge theory, and of several important ideas in modern physics, is an incorrect geometric theory of electromagnetism proposed by Hermann Weyl in 1918.  Weyl was inspired by Einstein's then-new theory of gravity.   The leading idea of Weyl's theory is that the fundamental laws of physics should be invariant under position-dependent scale transformations.   We implement those by changing the metric, according to
\begin{equation}\label{scalingMetric}
g^\prime_{\mu \nu} (x) ~=~ e^{2\sigma(x)}  \, g_{\mu \nu} (x)
\end{equation}
Note that this is a different operation than using an expanded coordinate system.  It implements real changes in size, or scale, or alternatively {\it gauge}, rather than changes in how we label things.

The term ``gauge symmetry'' is a relic of Weyl's original theory.

Weyl, in his original theory, postulated local scale symmetry.  That is, he postulated that one could change the size of objects independently at every point in space-time - and still get the same behavior!   To make that outrageous idea viable, he had to introduce a ``gauge'' connection field.   The gauge connection field tells us how much we must adjust our scale of length, or re-gauge our rulers, as we move from one point to another.   Weyl made the remarkable discovery, that this gauge connection field, in order to do its job of implementing local scale symmetry, must satisfy the Maxwell equations.   Dazzled by that apparent miracle, Weyl proposed to identify his ideal mathematical connection field with the real physical electromagnetic field.
 
Let me very briefly indicate how that proposal works.  

In order to have motion of particles or fields, we must have derivatives.  And if we want the derivatives to transform simply under gauge transformations, so that we can package them into invariant equations, we need to adjust them.   Specifically, if we consider the simplest non-trivial transformation law for a field $\phi(x)$,
\begin{equation}
\phi^\prime (x) ~=~ e^{-\sigma (x)} \phi(x)
\end{equation}
the derivative $\partial_\mu \phi$ picks up terms involving the derivative of $\sigma$, and transforms horribly.  To fix it, Weyl introduced a vector field $\omega_\mu$ and a covariant derivative operation $\nabla_\mu \phi = (\partial_\mu - \omega_\mu)\phi$.  Then we will have a good transformation law for the covariant derivative, {\it viz}.
\begin{equation}
\nabla^\prime_\mu \, \phi^\prime ~=~ (\nabla_\mu \phi)^\prime
\end{equation}
if $\omega$ transforms according to
\begin{equation}
\omega^\prime_\mu ~=~ \omega_\mu - \partial_\mu \sigma
\end{equation}
But this, mathematically, is identical to the standard gauge transformation law for electromagnetic potentials.  And, as we have seen, that gauge transformation law (plus special relativity) leads us pretty directly to the Maxwell equations.

Although Weyl's connection field is a necessary ingredient of local scale symmetry, it is not sufficient to ensure that symmetry.   Other properties of matter, such as the dimensions of a proton, give us objective scales of length that don't change as we move from point to point.   Einstein and others did not fail to notice this shortcoming of Weyl's theory.  Despite its visionary brilliance, Weyl's theory seemed destined for oblivion.  

The situation changed, however, with the emergence of quantum theory.  In that context electric charge associated with the phase of wave-functions, which we can regard as defining a circular one-dimensional property space, living on top of space-time \cite{weylPhase}.  
In 1929 Weyl exploited this new space, to revive his gauge theory in a modified form.  In the new theory, the local symmetry transformations are no longer space-time dependent changes in the scale of length  of space-time, but rather rotations in the new dimension, whose coordinate is dual to electric charge.   (Note that multiplication by a phase faithfully implements rotations of a circle.)     After that modification gauge symmetry leads to a satisfactory theory of electromagnetism, as I've sketched above.

\subsubsection{Scale Transformations as a Conceptual Tool}
  
There is a close connection between Weyl scaling Eqn.\,(\ref{scalingMetric}) with constant $\sigma$ and the scaling transformations  
\begin{equation}\label{flatScaling}
{x^\prime}^{\, \mu} ~=~ e^{-\sigma} x^\mu
\end{equation}
with {\it fixed}, flat metric that we often consider in high-energy and condensed matter physics.  If we perform the scaling Eqn.\,(\ref{flatScaling}) as a general coordinate transformation, and then follow it with the Weyl scaling Eqn.\,(\ref{scalingMetric}), we get a transformation that leaves the metric fixed.   So if general covariance is a valid symmetry, then the ``dimensional analysis'' re-scalings considered in high-energy and condensed matter physics are equivalent to constant Weyl transformations.   

In modern physics local Weyl transformations, though interesting, do not appear to be a good candidate to provide new fundamental symmetries, or new gauge interactions.   
There are major barriers to supposing that local scale invariance holds in Nature.  As we already observed, scale invariance would forbid basic properties we seem to need, like particle masses.   That difficulty might not be insuperable, since we have become accustomed to exploiting hidden (``spontaneously broken'') symmetries of equations that are not manifest in their stable solutions.   More difficult to finesse, it seems, is the fact that local scale symmetry is in tension with the  basic structure of interacting quantum field theory, due to the need for renormalization -- running of couplings.   This effect brings in distance dependence, even when the underlying classical theory is scale invariant.  

Remarkably, that difficulty may well prove to be a blessing in disguise.  For there are situations where that quantum effect, running of the couplings, is the main correction to scale symmetry.   Then we can gain insight from an appropriately corrected version of scale symmetry.  In the remainder of this lecture, I'll indicate how pursuing that line of thought enables us to quantify and refine an attractive proposal for the unification of force and substance.

\section{Unification of Quantum Numbers}

The ``standard model'' gauge theory $SU(3)\times SU(2) \times U(1)$  gives a successful, economical account of fundamental interactions based on far-reaching symmetry postulates, within the framework of relativity and quantum theory \cite{pdg}.   General relativity is also readily accommodated, using a minimal coupling procedure.   Since this theory is close to Nature's last word, we should take its remaining esthetic imperfections seriously.

One class of imperfections arises in the account of masses and mixings, where free parameters proliferate.  Unfortunately, there are few really compelling ideas that address this issue.  Even the most basic question: Why is there a three-fold ``family'' repetition of quarks and leptons? -- remains wide open.   (And if cosmic landscape ideas are right, it may be just an accident, with no deeper explanation.)

A second class of imperfections concerns the core of the standard model, that is the symmetry structure.  Here there is a compelling idea.  The product gauge symmetry structure $SU(3)\times SU(2) \times U(1)$ practically begs to be embedded into a larger, encompassing symmetry.   The electroweak theory, which breaks $SU(2) \times U(1)_Y \rightarrow U(1)_Q$, shows how the symmetry of fundamental equations can be hidden in their low-energy solutions, by the influence of cosmic fields or condensations (Higgs mechanism).   Slightly more elaborate versions of the same mechanism can implement
$SU(5) \rightarrow SU(3)\times SU(2)\times U(1)$ or $SO(10) \rightarrow SU(3)\times SU(2)\times U(1)$, as we'll presently discuss.   

An important test for the hypothetical expanded symmetries is whether they act naturally on the quarks and leptons.   Indeed, another ``imperfection'' of the symmetry of the standard model is that it classifies the quarks and leptons into several unrelated multiplets, even within one family.  If we allow for the right-handed neutrino $N$, needed to give a smooth theory of neutrino masses, there are six (if not, five).   Moreover the $U(1)_Y$ hypercharge quantum numbers we need to assign to those multiplets are funny fractions, determined phenomenologically.   

As we'll see, the extended symmetries -- especially $SO(10)$ -- do a brilliant job of organizing the fermion multiplets into a single one, and also of explaining the funny fractions.   

Although I won't develop it here, I should mention that one can constrain the hypercharges in an alternative way, by demanding anomaly cancellation.   That approach does not address the unification of couplings.  Nor does it explain the multiplet structure nearly so neatly: In particular, it does not predict the existence of the right-handed neutrino $N$, which plays a central role in the theory of neutrino masses.

\subsection{Standard Model Multiplets}

To get started, let's briefly review the multiplet structure of quarks and leptons in $SU(3)\times SU(2) \times U(1)$.  To keep things focussed I'll pretend there's just one family, and use the notations appropriate to the lightest quarks and leptons.   

Here's what we've got, in the usual presentation of the standard model:
\begin{eqnarray}
Q_L ~&=&~ \left(\begin{array}{ccc} u_L & u_L & u_L \\d_L & d_L & d_L \end{array}\right)^{\frac{1}{6}} \\
u_R ~&=&~ \left(\begin{array}{ccc}u_R & u_R & u_R \end{array}\right)^{\frac{2}{3}} \\
d_R ~&=&~ \left(\begin{array}{ccc} d_R & d_R & d_R \end{array}\right)^{-\frac{1}{3}} \\
L_L ~&=&~ \left(\begin{array}{c}\nu_L \\ e_L \end{array}\right)^{-\frac{1}{2}}  \\
e_R ~&=&~ \left( e_R \right)^{-1}
\end{eqnarray}
Here: 
\begin{itemize}
\item The $L, R$ subscripts denote left, respectively right, helicity.  
\item The threefold repetition of quark fields indicates three colors.  I dispensed with distinguishing indices.  I also let the $SU(3)$ act horizontally on the fundamental $\bf 3$ representation, for typographical clarity (whereas $SU(2)$ acts vertically). 
\item The numbers attached to the multiplets are their $U(1)_Y$ hypercharges.  These are given by the average electric charge of the particles in the multiplet, up to an overall normalization.
\end{itemize}

We can also write this in a less picturesque but more flexible way, using indices:
\begin{eqnarray}
Q_L ~&\equiv&~ Q_L^{\alpha a}{}_\frac{1}{6}   \\ 
u_R ~&\equiv&~ u_R^{\beta}{}_\frac{2}{3}  \\
d_R ~&\equiv&~ d_R^\gamma{}_{-\frac{1}{3}}  \\
L_L ~&\equiv&~  L_L^d{}_{-\frac{1}{2}} \\
e_R ~&\equiv&~ e_R{}_{-{1}} 
\end{eqnarray}
Here the Latin indices run from 1 to 2, and denote $SU(2)$ quantum numbers, while the Greek indices run from 1 to 3, and denote $SU(3)$ quantum numbers (colors).   

\pagebreak

For purposes of unification we will want to have fermions all of the same helicity.  We can do that by going to antiparticles.   In this way we arrive at
\begin{eqnarray}
Q ~&\equiv&~ Q^{\alpha a}{}_\frac{1}{6}   \\ 
\bar u ~&\equiv&~ {\bar u}_{\beta}{}_{-\frac{2}{3}} \\
\bar d ~&\equiv&~ {\bar d}_\gamma{}_{\frac{1}{3}}  \\
L ~&\equiv&~  L^d{}_{-\frac{1}{2}} \\
\bar e ~&\equiv&~ {\bar e}_{1} 
\end{eqnarray}
Here all the fields are understood to have left-handed helicity.   $\bar u$ and $\bar d$ are antitriplets $\bf {\bar 3}$ of $SU(3)$.   

\subsection{N} 

The theory of neutrino masses works most smoothly if we include an additional $SU(3)\times SU(2) \times U(1)$ singlet field $N$.  It (or rather, to be more precise, $\bar N$) can be considered as the right-handed neutrino.  

\subsection{Unification in $SU(5)$ \cite{georgiGlashow}}

It is easy to visualize how the action of $SU(5)$ can incorporate $SU(3)\times SU(2) \times U(1)$.  Indeed, a 5$\times$5 matrix contains  $3 \times$3 and $2\times 2$ blocks as follows:
\begin{equation}
\left(\begin{array}{ccccc}*** & *** & *** & 0 & 0 \\ **** & *** & *** & 0 & 0 \\ **** & *** & *** & 0 & 0 \\0 & 0 & 0 & ** & ** \\0 & 0 & 0 & ** & **\end{array}\right) 
\end{equation}
Now we need to implement the ``$S$'' part of $SU(n)$ -- getting unit determinants.  Within $SU(5)$, but not within $SU(3)\times SU(2)$, we have the $U(1)$ subgroup
\begin{equation}
\left(\begin{array}{ccccc}e^{i2\theta} & 0 & 0 & 0 & 0 \\0 & e^{i2\theta} & 0 & 0 & 0 \\0 & 0 & e^{i2\theta} & 0 & 0 \\0 & 0 & 0 & e^{-i3\theta} & 0 \\0 & 0 & 0 & 0 & e^{-i3\theta}\end{array}\right)
\end{equation}
This identifies, essentially uniquely, $SU(3)\times SU(2) \times U(1)$ as a subgroup of $SU(5)$.  

Now let us see how the fermions might fit in this picture.   There aren't an enormous number of them -- 15 or 16, depending on whether you include $N$ -- relative to 5.  So we must look to low-dimensional representations, specifically to vectors (${\bf 5}, {\bf \bar 5}$) or two-index antisymmetric tensors (${\bf 10}, {\bf \bar {10}}$).   In making the initial comparison, we will allow the normalization of the hypercharge to float, using
\begin{equation}
Y ~=~ c \, \left(\begin{array}{ccccc}-2 & 0 & 0 & 0 & 0 \\0 & -2 & 0 & 0 & 0 \\0 & 0 & -2 & 0 & 0 \\0 & 0 & 0 & 3 & 0 \\0 & 0 & 0 & 0 & 3\end{array}\right)
\end{equation}
with $c$ a free parameter.   The value of $c$ is an important element in the quantitative aspect of unification, as we'll see.   

A little experimentation reveals that we get the right representation and charge spectrum by choosing the representations ${\bf \bar 5} \oplus {\bf 10}$ and $c = \frac{1}{6}$.  Let's spell that out.   

First, a simple but vital observation: In $SU(2)$, doublets and antidoublets are equivalent. Indeed, we can trade one for the other using the (invariant) antisymmetric symbol $\epsilon$, in the style
\begin{eqnarray}
r^a ~&\leftrightarrow&~ \epsilon^{ab} r_b \nonumber \\
r_b ~&\leftrightarrow&~ -\epsilon_{bc} r^c
\end{eqnarray}


The first three components of ${\bf \bar 5}$ make a color antitriplet weak singlet, and the hypercharge is $-(-2)c = \frac{1}{3}$.  That matches the quantum numbers of $\bar d$.   The last two components make a color singlet weak doublet with hypercharge $-3c = -\frac{1}{2}$.  (Recall that $SU(2)$ doublets and antidoublets are equivalent.)  That matches the quantum numbers of the lepton multiplet $L$.  

The components of $\bf 10$ with two early indices -- that is, $T^{ab}$ with $a,b = 1, 2, 3$ -- make a color {\it antitriplet\/} weak singlet with hypercharge $((-2) + (-2)) c = - \frac{2}{3}$.  That matches the quantum numbers of $\bar u$.  The components with one early and one late index make a color triplet weak doublet with hypercharge $\frac{1}{6}$.  That matches the quantum numbers of the quark multiplet $Q$.  Finally, the component $T^{45}$ with two late indices defines a color singlet weak singlet with hypercharge $1$.  That matches the quantum numbers of  $\bar e$.  

Thus all the quarks and leptons have been accommodated neatly, with no loose ends, and the mishmash of funny hypercharges has been rationalized into the one number $c$, with the consistent value
\begin{equation}
c ~=~ \frac{1}{6}
\end{equation}
$N$ can be brought in as an $SU(5)$ singlet, since it is an $SU(3)\times SU(2) \times U(1)$ singlet, with hypercharge 0. 

\subsection{Unification in $SO(10)$ \cite{georgiGlashow}}

The spinor representation is central to $SO(10)$ unification, so let me recall (or reveal) how that works.  

It is convenient to begin with the Clifford algebra
\begin{equation}
\{ \gamma_j, \gamma_k \} ~=~ 2 \delta_{jk}
\end{equation}
where $j, k$ run from 1 to 10.   This algebra can be realized in a useful form, very familiar to physicists, by defining
\begin{eqnarray}\label{fermionClifford}
a_j ~&=&~ \frac{1}{2} (\gamma_{2j-1} - i \gamma_{2j} ) \nonumber \\
a_j^{\, \dagger} ~&=&~ \frac{1}{2} (\gamma_{2j-1} + i \gamma_{2j}) 
\end{eqnarray}
for the $a$s behave as a set of five fermion creation and destruction operators:
\begin{eqnarray}
\{ a_j, a_k^{\, \dagger} \} ~&=&~ 2 \delta_{jk} \nonumber \\
\{ a_j, a_k \} ~&=&~ 0 \nonumber \\
\{ a_j^{\, \dagger}, a_k^{\, \dagger} \} ~&=&~ 0
\end{eqnarray}
We can of course also invert the definition Eqn.\,(\ref{fermionClifford}), to get the Clifford algebra from fermions. 


The connection of the Clifford algebra to $SO(10)$ is that the commutators
\begin{equation}\label{rotationsFromClifford}
R_{jk} ~=~ \frac{1}{4} [ \gamma_j, \gamma_k ] 
\end{equation}
satisfy, by virtue of the Clifford algebra, the commutation relations of infinitesimal rotations.  If we define $R_{jk}$ to represent an infinitesimal rotation in the $jk$ plane, we get the same Lie algebra as we get from Eqn.\,(\ref{rotationsFromClifford}).  I will leave the detailed verification as an exercise, but here only check the corresponding logic for $SO(3)$.  In that case we can realize the Clifford algebra using Pauli spin matrices
\begin{equation}
\gamma_j ~=~ \sigma_j
\end{equation}
Our recipe gives, for example 
\begin{equation}
R_{12} ~=~ \frac{1}{4} [ \sigma_1, \sigma_2 ] ~=~ i \frac{\sigma_3}{2}
\end{equation}
which indeed is the standard generator of rotations in the 12 plane -- i.e., around the 3 axis -- in the theory of spin $\frac{1}{2}$.  

Going back to the fermion representation, we see that the $SO(10)$ generators generally contain pieces of the type $a_j a_k$ and their Hermitean conjugates, that change fermion number by two units.   On the other hand these cancel in the rotations involving $a$s with a fixed index, i.e. $R_{12}, R_{34},R_{56},$ $ R_{78}, R_{9(10)}$.  Indeed for example
\begin{equation}
R_{12} ~=~ i (a_1^{\, \dagger} a_1 - \frac{1}{2} )
\end{equation}
so these essentially count the occupancy of the different fermion states.  

More generally, we find that the number-changing terms cancel in combinations like
\begin{equation}
R_{13} + R_{24} ~\propto~ [ a_1 + a_1^{\, \dagger} , a_2 + a_2^{\, \dagger} ] \, - \, [ a_1 - a_1^{\, \dagger} , a_2 - a_2^{\, \dagger} ]
\end{equation}
and $R_{14} - R_{23}$, to give us the possibility of exchanging fermions of types 1 and 2.  In this way, we generate the full $U(5)$ symmetry of 5 distinguishable but equivalent free elementary fermions.   The $U(1)$, which simply counts total number, is of course essentially $R_{12} + R_{34} + R_{56} + R_{78} + R_{9(10)}$.  We can, alternatively, define our $U(5)$ as the centralizer of this transformation, i.e. the number-conserving transformations within $SO(10)$.   

The minimal realization of fermions gives us a $2^5 = 32$ dimensional state space on which $SO(10)$ acts.  That action is not quite irreducible, however, because $SO(10)$ conserves fermion number modulo 2.  So we have two invariant 16 dimensional state spaces wherein the fermion number is $(0, 2, 4)$ or $(1, 3, 5)$ respectively.  These turn out to be irreducible (and inequivalent).  For unification purposes, with our realization of $SU(5)$ as transformations among elementary fermions, we want the former.
Indeed, since the fermion label transforms as a vector under $SU(5)$, and the states are antisymmetric in that label (fermionic), $(0, 2, 4)$ will give us a singlet, an antisymmetric 2-index tensor ${\bf 10}$, and an anti-vector $\bf \bar 5$.  As we saw previously, that is just what we need to define the observed quarks and leptons.   

In this construction all the directly observed fermions are combined into a single irreducible representation, together with $N$.   $N$ has been promoted from a desirable option to a necessary feature.  Now it forms part of the operating system.

\section{Unification of Coupling Strengths}

In the preceding section we've seen how the messy multiplet structure of quarks and leptons in the standard model, including their hypercharges, comes to look much nicer when viewed within the context of $SU(5)$ or especially $SO(10)$ unification.    

Those higher symmetries realize the separate $SU(3)\times SU(2)\times U(1)$ gauge symmetries of the standard model as different aspects of encompassing symmetry.   The glory of local (gauge) symmetry, however, is that it controls not only bookkeeping, but also dynamics.   For a simple (in the technical sense) gauge group  such as $SU(5)$ or $SO(10)$, symmetry dictates all the couplings of the gauge bosons, up to a single overall coupling constant.    

Thus unification predicts relationships among the strong, weak, and hypercharge couplings.  Basically -- up to the group-theoretic task of normalization --  it predicts that the three couplings for $SU(3)\times SU(2)\times U(1)$ must be equal.   
As observed, of course, they are not.  But the two great dynamical lessons of the standard model -- namely symmetry breaking through field condensation (Higgs mechanism), and running of couplings (asymptotic freedom) -- suggest a way out \cite{gqw}.  We can imagine that the symmetry breaking $G \rightarrow SU(3)\times SU(2) \times U(1)$ occurs through a big condensation, at a high mass scale.   In the symmetric theory, appropriate to the description of processes at large mass scales, there was only one unified coupling.   But we make our observations at a much lower mass scale.  To get to the unified coupling, we must evolve the observed couplings up to high energy, taking into account vacuum polarization.  Note that throughout that evolution the unified symmetry is violated, so the three $SU(3)\times SU(2) \times U(1)$ couplings evolve differently.   

Before entering the details, let us pause to consider a soft-focus view of what we can expect from this sort of calculation.  Our input will be the observed couplings, plus some hypothesis $\cal H$ about the spectrum of virtual particles we need to include in the vacuum polarization.   Our output should be the unified coupling strength, and the scale of unification.   For any given $\cal H$, we have three inputs -- the observed couplings -- and two outputs -- the scale and coupling at unification.  So there will be a consistency condition.  If the calculation works, we will have reduced the number of free parameters in the core of the standard model by one, from three to two.   

There are additional {\it physical\/} consistency conditions, concerning the value of the unification scale, which are quite significant.  Also, in case of success, we will need to discuss the plausibility of our hypothesis $\cal H$.  We'll return to these important points later, after we've done the central calculation.

\subsection{Normalization of Hypercharge}

The value of nonabelian couplings can be fixed absolutely, because the generators obey nonlinear commutation relations.   It is common practice, for $SU(n)$ or $SO(n)$ groups, to choose the coupling constant to multiply, in the fundamental representation, generators the trace of whose square $\frac{1}{2}$.  Thus for $SU(2)$ we have the covariant derivative
\begin{equation}
\nabla_\mu ~=~ \partial_\mu + i g_2 \frac{\sigma_a}{2} A^a_\mu 
\end{equation}
for isospinors, and so forth.   (Of course, we understand here that the $A^a$ appear in a Maxwell-like (Yang-Mills) action, in such a way that the plane wave ``generalized photons'' are canonically normalized.)   

That fixes the normalizations for $g_3$, $g_2$, the couplings associated with $SU(3)$ and $SU(2)$ respectively, and also the normalization of $g_5$ in $SU(5)$\footnote{There is no need, here, to discuss $SO(10)$ separately.}.    

The normalization of the hypercharge generator in $SU(5)$ is also thereby fixed, but its numerical relationship to the conventional normalization of the hypercharge of $U(1)_Y$, as it appears in electroweak phenomenology, requires discussion.   

We have seen that in $SU(5)$ the fundamental representation (actually its conjugate, but that makes no difference here) is implemented on $(\bar d, \bar d , \bar d, L)$.   The trace of the square of the hypercharge generator, times the square of the coupling constant, acting on this is therefore $g_5^{\, 2}/2$.   On the other hand, if we evaluate the same thing in the standard electroweak notation, we get 
\begin{equation}
(g^\prime)^2 ( 3 \times (\frac{1}{3})^2 \, + \, 2 \times (\frac{1}{2})^2 ) ~=~ (g^\prime)^2 \, \times \, \frac{5}{6}
\end{equation}
Equating these two evaluations of the same thing, we have
\begin{eqnarray}
g_5^{\, 2} \, \times \, \frac{1}{2} ~&=&~  (g^\prime)^2 \, \times \, \frac{5}{6} \nonumber \\
g_5^{\, 2} \, \times \, \frac{3}{5}  ~&=&~  (g^\prime)^2
\end{eqnarray}

It will be convenient, for later purposes, to express this result as the definition
\begin{equation}
g_1^{\, 2} ~\equiv~ \, \frac{5}{3} (g^\prime )^2
\end{equation}
In the unified theory it is $g_1$ that should evolve to become equal to $g_2$ and $g_3$ (and $g_5$).   

Before including any running of the couplings, we get the two ``predictions''
\begin{eqnarray}
g_3^{\, 2} ~&=&~ g_2^{\, 2} ~(\ = g_5^{\, 2}\, ) \nonumber \\
\sin^2 \, \theta_W ~&\equiv&~ \frac{(g^\prime)^2}{g_2^2 + (g^\prime)^2} ~=~ \frac{\frac{3}{5}}{1 + \frac{3}{5}} ~=~ \frac{3}{8}
\end{eqnarray}
They are way off.

\subsection{Structure of Coupling Renormalization}

Each of the couplings is affected by vacuum polarization.  Thus the values observed in an experiment will depend on the distance, or equivalently the energy and momentum, characteristic of the measurement process. 

The logarithmically divergent (before regularization) terms are proportional, in lowest order, to the cube of the coupling\footnote{This emerges most clearly if we consider the gluon self-coupling.  When we consider couplings of gluons to fermions, there appears to be cross-talk between the different interactions.   These cross terms turn out not to contribute, due to Ward's identity, which is a diagrammatic manifestation of gauge symmetry.  They'd better not, because the renormalized non-abelian coupling should be universal -- the same for gluons, fermions, scalars, ghosts, ... .}.   So we find equations for the running couplings of the form 
\begin{equation}\label{rgFirstForm}
\frac{d \bar g_j(Q)}{d \ln Q} ~\approx~ b_j {\bar g_j}^{\, 3}
\end{equation}
where $b$ is a number that depends on the spectrum of virtual particles that contribute.

To expose the logic of coupling unification, it is helpful to re-write Eqn.\,(\ref{rgFirstForm}) as 
\begin{equation}
\frac{d\, 1/{\bar g_j}^{\, 2}}{d \ln Q} ~=~ -\frac{2}{{\bar g_j}^{\, 3}} \frac{d \bar g_j(Q)}{d \ln Q} ~=~ -2b_j
\end{equation}
with the solution
\begin{equation}\label{rgSolution}
\frac{1}{{{\bar g_j}(Q)}^{\, 2}} ~=~ - 2 b_j \, \ln \frac{Q}{Q_0} \, + \, \frac {1}{\bar g_j(Q_0)^{\, 2}}
\end{equation}
Here we take $Q_0$ to be an accessible laboratory scale, where we do empirical measurements of the  ${\bar g_j(Q)^{\, 2}}$.  

Now the unification condition is that for some $Q$ the effective couplings ${\bar g}_3^{\, 2} (Q), {\bar g}_2^{\, 2} (Q),{\bar g}_2^{\, 2} (Q)$ become equal to a common value, call it $g_5(Q)$.  By subtracting the solutions Eqn.\,(\ref{rgSolution}) for $j=2,3$ we derive an equation determining the unification scale:
\begin{equation}\label{scaleEquation}
2 \, (b_3 - b_2) \ln \frac{Q}{Q_0} ~=~ \frac {1}{\bar g_3(Q_0)^{\, 2}} \, - \, \frac {1}{\bar g_2(Q_0)^{\, 2}}
\end{equation}
Since we must derive the same scale from other pairs of couplings, we have the consistency condition
\begin{equation}\label{couplingEquation}
\frac{b_3 - b_2}{b_2 - b_1} ~=~ \frac{\frac {1}{\bar g_3(Q_0)^{\, 2}} \, - \, \frac {1}{\bar g_2(Q_0)^{\, 2}}} { \frac {1}{\bar g_2(Q_0)^{\, 2}} \, - \, \frac {1}{\bar g_1(Q_0)^{\, 2}} }
\end{equation}
This is the anticipated prediction, supplying a numerical relation among the observed couplings.   

Of course, once we have the unification scale, we can go back to determine the value of the unified coupling using Eqn.\,(\ref{rgSolution}).  

\subsection{Numerics of Couplings Renormalization}

\subsubsection{Renormalization Group Coefficients}

The renormalization group coefficients $b_j$ can be calculated perturbatively, for any combination of spin 0, $\frac{1}{2}$, and 1 (gauge) fields.   The result is \cite{grossWilczek, politzer}
\begin{equation}\label{rgCoefficients}
16\pi^2 \, b ~=~ - \frac{11}{3} C_A + \frac{4}{3} T(R_{\frac{1}{2}}) + \frac{2}{3} T(R_0)
\end{equation}
Here $C_A$ is the value of the Casimir operator for the adjoint representation of the gauge group in question.  Explicitly, we have 
\begin{equation}
C_A (SU(n)) ~=~ n 
\end{equation}
$T(R_{\frac{1}{2}})$ is the trace of the square of a normalized generator acting on the spin-$\frac{1}{2}$ fields in the theory, which may of course include several multiplets. We will only need the basic defining normalization $T = \frac{1}{2}$ for the fundamental representation, and, when we come to consider supersymmetry, $T = C = n$ for the adjoint of $SU(n)$.   The coefficient $\frac{4}{3}$ for fermions holds for complex Dirac fermions.  For Weyl fermions (i.e., fermion fields with definite helicity) we get half that, as we do for Majorana (real) fermions.   $T(R_0)$, unsurprisingly, is the trace of the square of a normalized generator acting on the spin-$0$ fields in the theory, which may of course include several multiplets.  The coefficient $\frac{1}{3}$ holds for complex scalars; for real scalars we would get half that (but that case will not arise below).   

\subsubsection{Minimal Extrapolation}

Taking the particles of the standard model, but allowing (why not?) for $n_f$ families and $n_s$ Higgs doublets, we have
\begin{equation}
16\pi^2 \, b_3 ~=~ - 11 \, + \, \frac{4}{3} \, n_f
\end{equation}
since there are two fundamentals of Dirac fermions per family.   Similarly
\begin{equation}
16\pi^2 \, b_2 ~=~ - \frac{22}{3} \, + \, \frac{4}{3} \, n_f \, + \, \frac{1}{3} n_s
\end{equation}
(four fundamentals of Weyl fermions!) and finally 
\begin{equation}
16\pi^2 \, b_1 ~=~ \frac{4}{3} \, n_f \, + \, \frac{1}{5} n_s
\end{equation}
The $b_1$ equation can be obtained painlessly by noting that the contribution of complete families must respect the $SU(5)$ symmetry -- that remark governs the fermions directly, while the Higgs particle should be fleshed out with a color triplet that isn't there, so its contribution is reduced by the factor
\begin{equation}
\frac{2 \times (\frac{1}{2})^2 }{ 2 \times (\frac{1}{2})^2 \, + \, 3 \times (\frac{1}{3})^2 } ~=~ \frac{3}{5}
\end{equation}

If we run the couplings using these coefficients, with $n_f = 3$, $n_s =1$, we get an unsatisfactory result 
\cite{raby}.   

\subsubsection{Extrapolation With Supersymmetry \cite{drw}}

Supersymmetry is too big, and too technical, a subject to develop here for a general audience of physicists.  I will only insert two brief, broad motivating comments, which I find irresistible.   
\begin{itemize}
\item Wave-particle duality blurred the contrast between force, epitomized by electromagnetism and emergent light, and substance, epitomized by electrons.   At the level of single-particle quantum mechanics, photons and electrons fall comfortably into a single framework.  This changes, however, at the level of multi-particle quantum mechanics, where the contrasting quantum statistics of force (bosons) and substance (fermions) marks a sharp dichotomy of matter.  Supersymmetry, by allowing transformations that exchange bosons and fermions, re-establishes unity.
\item Our unification, described above, brings together all the forces, on the one hand, and all the substances, on the other\footnote{Close readers will appreciate that I've allowed myself some poetic license here.}.   But even were it consummated, it would leave us with two separate things: force and substance.   Supersymmetry bring those together.
\end{itemize}

To implement low-energy supersymmetry, in a minimal fashion, we must expand the standard model in several ways:
\begin{enumerate}
\item We have spin-$\frac{1}{2}$ Majorana fermion partners of the gauge fields, in the adjoint representation.  Grouping their contribution with the gluons, has the effect of changing the $- \frac{11}{3}$ in Eqn.\,(\ref{rgCoefficients}) to $-3$.  
\item For each chiral fermion in the theory, we get a complex scalar superpartner.  Grouping their contribution with the fermions, this has the effect of changing the $\frac{4}{3}$ in Eqn.\,(\ref{rgCoefficients}) to $2$ -- still understanding, of course, that we halve this for chiral fermions!
\item Conversely, for each ordinary Higgs doublet we need a chiral fermion with the same quantum numbers.  This changes the $\frac{1}{3}$ in Eqn.\,(\ref{rgCoefficients}) to $1$
\item Supersymmetry requires, at a minimum, $n_s = 2$ Higgs doublets.
\end{enumerate}

Putting all this together, we find that when the contribution of virtual particles required by supersymmetry is included, we have
\begin{eqnarray}
16\pi^2 \, b_3 ~&=&~ - 9 \, + \, 2 \, n_f \,  \nonumber \\
16\pi^2 \, b_2 ~&=&~ - 6 \, + \, 2 \, n_f \, + \, \frac{1}{2} n_s \nonumber \\
16\pi^2 \, b_1 ~&=&~  \ \ \ 0 \, + \, 2 \, n_f \, + \, \frac{3}{10} n_s 
\end{eqnarray}

If we run the couplings using these coefficients, with $n_f = 3$, $n_s =2$, we get a much more satisfactory result \cite{raby}.  The essence of the matter is that our predictive equation, Eqn.\,(\ref{couplingEquation}), is now well satisfed.   The unification scale is computed to be $\approx 2 \times 10^{16}$ GeV.   

As a measure of the delicacy and resolving power of the calculation, let us note that taking $n_s = 4$ leads to a 15\% error in the prediction of the Weinberg angle, if we use strong-weak unification to fix the scale.   

\subsubsection{Observations on the Generality} 

{\it An important general observation}: To the order we have been working, complete $SU(5)$ multiplets affect neither the predicted relation among observed couplings, nor the predicted scale of unification \cite{drw}!   That striking conclusion follows because complete multiplets contribute equally to $b_1, b_2$, and $b_3$, and in Eqns.\,(\ref{couplingEquation}, \ref{scaleEquation}) only differences among those coefficients occur.   So our successful ``minimal supersymmetry'' hypothesis is not so special as might appear at first sight.   

For example, one need not postulate a complete desert (apart from supersymmetry) in the mass spectrum between current observations and the unification scale.  One can populate it with any number of singlets, or with a modest number of complete families.  It is only broken families that are worrisome.   We can also allow, within supersymmetry, the masses of squarks and sleptons -- the partners of quarks and leptons -- to float up to a high scale together, since they form complete $SU(5)$ multiplets.  In principle, there could even be different large masses for the different families of superpartners.   

The partners of gauge bosons, on the other hand, {\it do not\/} form a complete $SU(5)$ multiplet.  Their masses cannot be allowed to float very high before ruining the success of our calculation.   A recent estimate \cite{bourilkov} suggests $M \sim 2$ TeV is the preferred scale.

Raising the squark and slepton masses is an attractive option phenomenologically, because it relieves difficulties with proton decay and flavor violation -- processes that otherwise tend to be over-predicted in supersymmetric models.  These possibilities have been explored in speculative phenomenology, first under the epithet ``focus point'' and more recently also as ``split'' and ``mini-split'' supersymmetry.   (On the other hand, we may not want to raise those masses $> 10$ TeV, as this leads to difficulties with another attractive consequence of unified theories, namely their quantitative explanation of the mass ratio $m_b/m_\tau$ \cite{raby}.)

Effects of complete multiplets will show up in more accurate calculations, taken to higher order.  They also affect the value of the unified coupling.   By increasing the value of $b_j$  in Eqn.\,(\ref{rgSolution}), they make the unified coupling larger.   We probably don't want to have too many of them, therefore.


\subsection{Loose Ends}

Above I have outlined the lowest-order calculation.  There is a vast technical literature on corrections, both those due to additional couplings and those due to masses \cite{raby}.   I think it is fair to say that the situation is generally satisfactory, although unfortunately there are many more potentially relevant parameters than experimental constraints, when one attempts precision work.   

The Higgs doublet of the standard model, or the two Higgs doublets of its supersymmetric extension, do not fill out unified multiplets.  Indeed, there are powerful bounds on the mass of the possible color triplet partners, since they make it difficult to maintain baryon number conservation as a good approximation.    There are several ideas to address this doublet-triplet splitting embarrassment, but no consensus on which (if any) is correct.

\section{Prospect}

\subsection{Significance of the Scale}

\subsubsection{Relation to Planck Scale}

The Planck energy
\begin{equation}
{\cal E}_{\rm Planck} ~=~ \sqrt{ \frac {\hbar c^5}{8 \pi G_N} } ~\approx~ 2.4 \times 10^{18} \, {\rm GeV}
\end{equation}
is another famous energy scale that can be constructed from fundamental constants\footnote{We have quoted the so-called rationalized Planck scale, including the factor $8\pi$ that naturally appears with $G_N$ in the Lagrangian of general relativity.}.  Here the construction is simple dimensional analysis, based on Newton's gravitational constant $G_N$ together with $\hbar, c$.    On the face of it, Planck units set the scale for effects of quantum gravity; thus when we consider basic (technically: ``hard'') processes whose typical energies are of order $E$, we expect gravitational effects of order $(E/{\cal E}_{\rm Planck})^2$.   

Our scale ${\cal E}_{\rm unification} \approx 2 \times 10^{16}$ GeV is significantly, but not grotesquely, smaller than the Planck scale.  This means that at the unification scale the strength of gravity, heuristically and semi-quantitatively, is of order 
\begin{equation}
({\cal E}_{\rm unification}/{\cal E}_{\rm Planck})^2 ~\sim~ 10^{-4}
\end{equation}
to be compared with the strength $g_5^{\, 2} /4\pi ~\sim~ 10^{-2}$ for the other interactions.   The relative smallness of gravity, thus estimated, which is of course further accentuated at lower energies, suggests that our neglect of quantum gravity in the preceding calculations may be justified.  

On the other hand, it seems to me remarkable that the comparison comes so close.  A classic challenge in fundamental physics is to understand the grotesque smallness of the observed force of gravity, compared to other interactions, as it operates between fundamental particles .   Famously, the gravitational interaction is $\sim 10^{42}$ times smaller than 
any of the other forces.   Again, however, proper comparison requires that we specify the energy scale at which the comparison is made.   Since the strength of gravity, in general relativity, depends on energy directly, it appears hugely enhanced when observed with high-energy probes.  At the scale of unification ${\cal E}_{\rm unified} \sim 2\times 10^{16} {\rm GeV}$ the discrepant factor $10^{42}$ is reduced to $\sim 10^4$, or even a bit less.  While this does not meet the challenge fully, it is a big step in the right direction.

\subsubsection{Neutrino Masses and Proton Decay}

By expanding our theory, unification along the lines we have been discussing brings in additional interactions.  Since the unified multiplets combine particles that normally (i.e., with the standard model itself!) don't transform into one another, we find new processes of transformation.   The two classic predictions for ``beyond the standard model'' interactions are small neutrino masses, leading to neutrino oscillations, and proton decay.   The first has been been vindicated; the second not yet.   In both cases, the large scale ${\cal E}_{\rm unification}$ is crucial for explaining the smallness of the new effects.  For an authoritative review of these and other aspects of unification, emphasizing the phenomenological issues, with many further references, see \cite{raby}.

\subsection{Conclusion}

The unification of quark and lepton quantum numbers in $SU(5)$, and especially $SO(10)$, is smooth and strikingly beautiful.  The unification of coupling strengths fails quantitatively if one makes a minimal extrapolation of the standard model, but under the hypothesis of low-energy supersymmetry its success is likewise smooth and strikingly beautiful.  The unification of coupling strengths brings a new scale into physics, which has several attractive features including, notably, a big step toward unification with gravity.    The discovery (or not) of some superpartners at the LHC will bring this line of thought to fulfillment (or not).   If it does, we shall have not only unified the different forces with   one another, and the different substances with one another -- but also unified force and substance, fulfilling Maxwell's vision.


\bigskip

{\it Acknowledgement}: 
This work is supported by the U.S. Department of Energy under contract No. DE-FG02-05ER41360.

\end{document}